\documentclass{sf2a-conf2021}
\usepackage{graphicx}
\usepackage{hyperref}
\usepackage[]{natbib}  
\usepackage{epstopdf}

\def\BibTeX{{\rm B\kern-.05em{\sc i\kern-.025em b}\kern-.08em
    T\kern-.1667em\lower.7ex\hbox{E}\kern-.125emX}}
\bibpunct{(}{)}{;}{a}{}{,}  

\begin{document}

\TitreGlobal{SF2A 2021}


\title{X-IFU/ATHENA VIEW OF THE MOST DISTANT GALAXY CLUSTERS IN THE UNIVERSE}

\runningtitle{X-IFU view of distant galaxy clusters}

\author{F. Castellani}\address{IRAP, Univ. de Toulouse, CNRS, UPS, CNES, Toulouse, France}
\author{N. Clerc$^1$}
\author{E. Pointecouteau$^1$}
\author{Y. Bahé}\address{Leiden Observatory, Leiden University, Leiden, the Netherlands}
\author{F. Pajot$^1$}



\setcounter{page}{237}


\maketitle


\begin{abstract}
The X-ray Integral Field Unit (X-IFU) on-board the second large ESA mission “Athena” will be a high spatial (5”) and spectral (2.5eV) resolution X-ray imaging spectrometer, operating in the 0.2-12 keV energy band. It will address the science question of the assembly and evolution through cosmic time of the largest halos of matter in the Universe, groups and clusters of galaxies. To this end, we present an on-going feasibility study to demonstrate the X-IFU capabilities to unveil the physics of massive halos at their epoch of formation. Starting from a distant (z=2) group of galaxies ($M_{500}$ = 7 $10^{13}$ $M_\odot/h$) extracted from the HYDRANGEA cosmological and hydrodynamical numerical simulations, we perform an end-to-end simulation of X-IFU observations. From the reconstruction of the global, 1D and 2D quantities, we plan to investigate the various X-IFU science cases for clusters of galaxies, such as the chemical enrichment of the intra-cluster medium (ICM), the dynamical assembly of groups and clusters and the impact of feedback from galaxy and super-massive black hole evolution.
\end{abstract}



\section{Introduction}

Athena is the ESA second large mission of Cosmic Vision program, dedicated to study the Hot and Energetic Universe \citep{nandra}. On-board Athena, the X-ray Integral Field Unit \citep{barret} (X-IFU) is a cryogenic imaging spectrometer composed by an array of 3168 superconducting Transition Edge Sensors \citep{smith} operated at 90mK. It will provide high spatial resolution (5” within an FoV of 5’ equivalent diameter) and high spectral resolution (2.5eV FWHM up to 7 keV) observations in the 0.2-12 keV energy band. The X-IFU is reaching the end of its preliminary definition phase (phase B). In order to assess the specifications of its performance, we put in place feasibility studies of the core science objectives of Athena to be carried out by X-IFU \citep{pointecouteau,croston}.
  
\section{Mock X-IFU observations}

\begin{figure}[ht!]
 \centering
 \includegraphics[width=0.33\textwidth,clip]{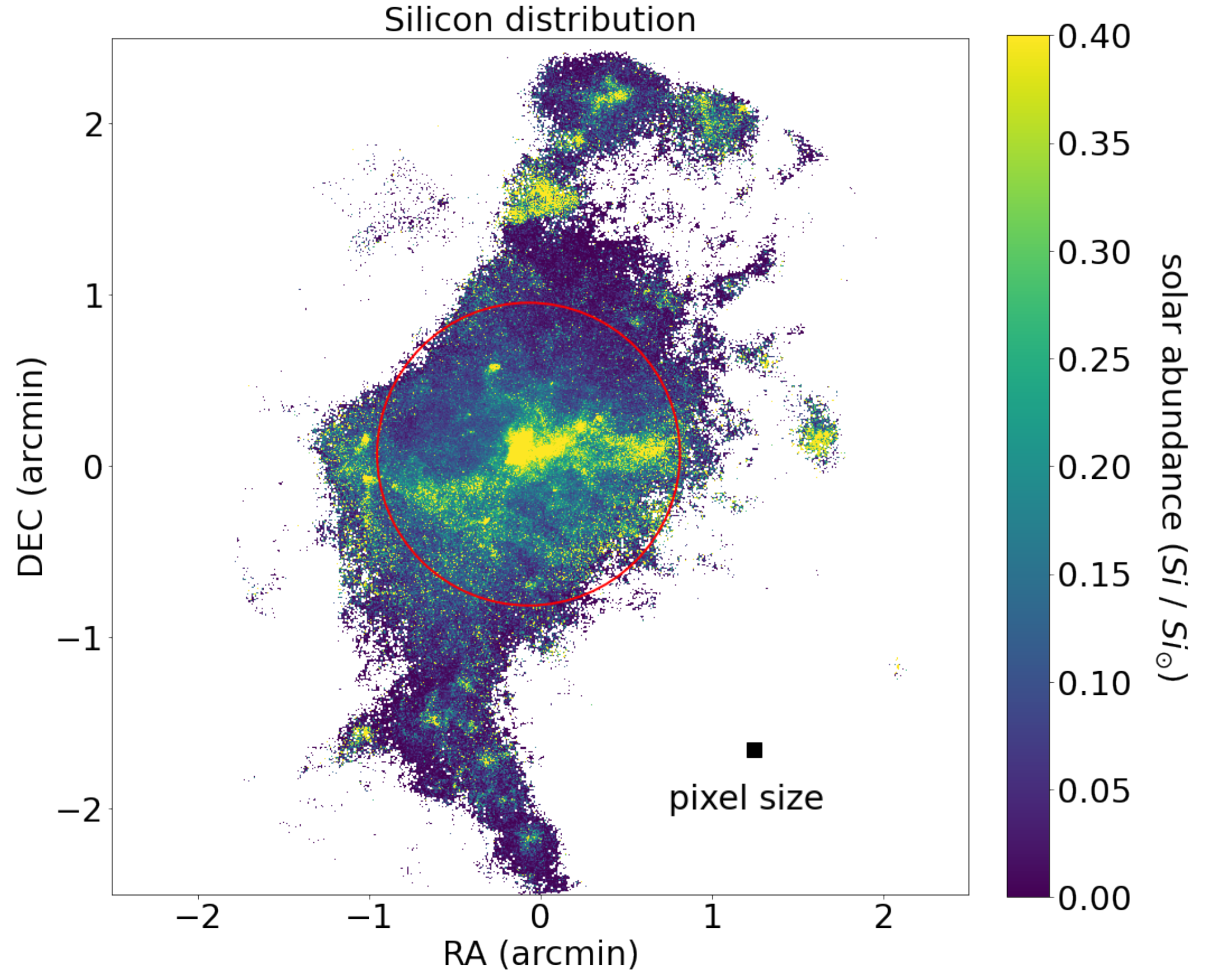}%
 \includegraphics[width=0.33\textwidth,clip]{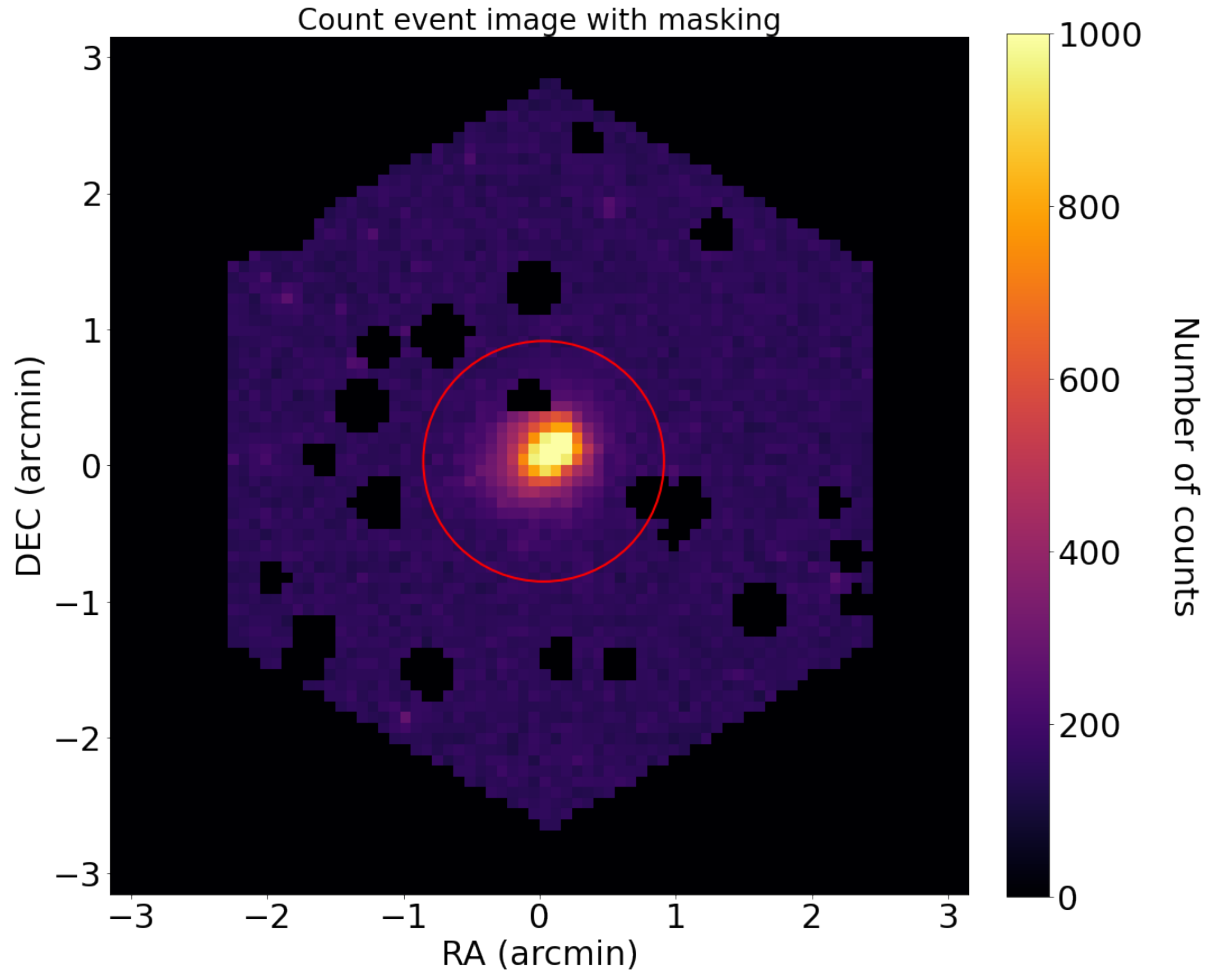}
 \includegraphics[width=0.33\textwidth,clip]{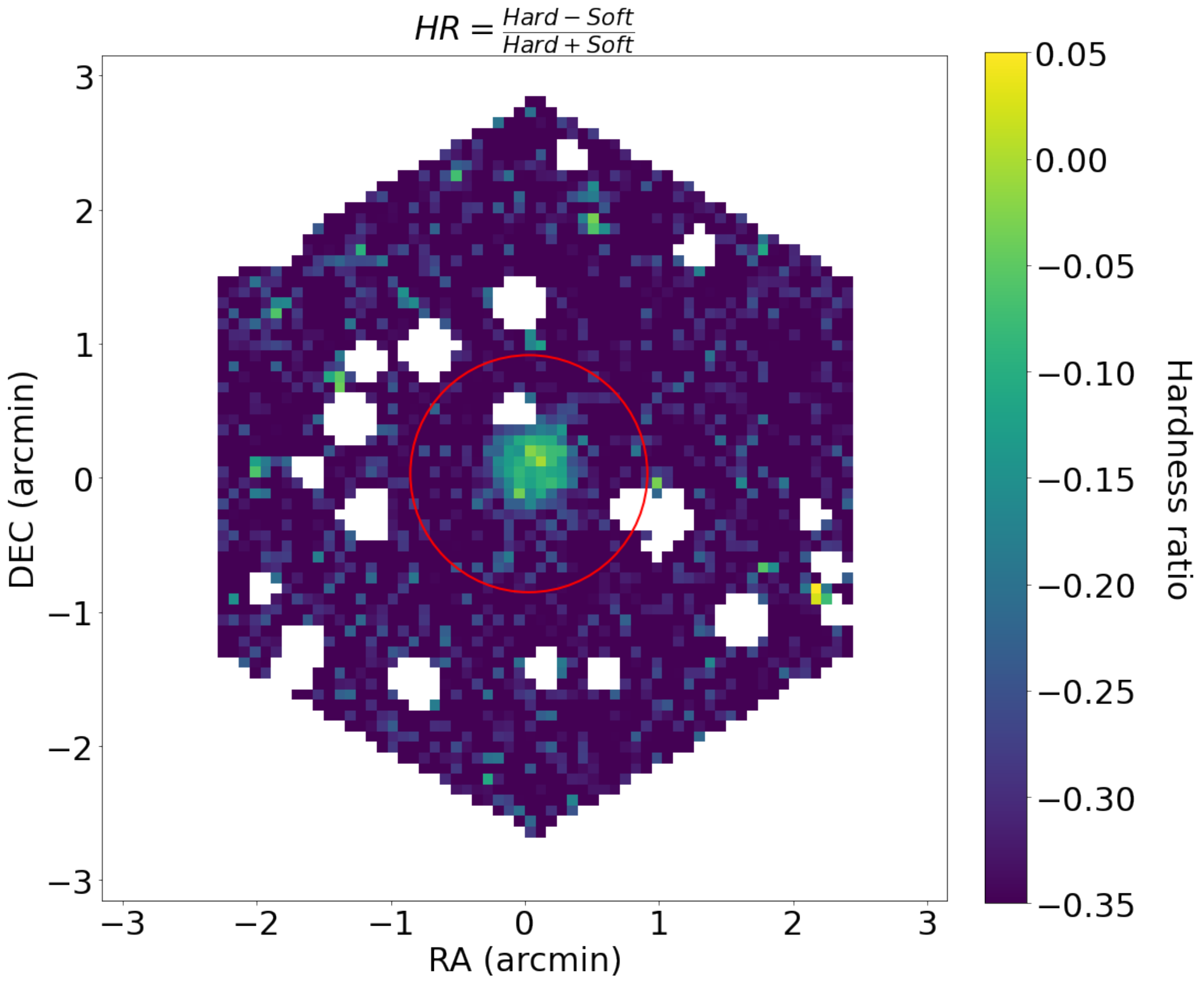}
  \caption{{\bf Left:} Emission-measure weighted map of the silicon abundances derived from the input simulated cluster gas particles. {\bf Middle:} Count image of our mock X-IFU observation including astrophysical and instrumental background (detected point sources are masked). {\bf Right:} Hardness ratio map from the count images derived in the 0.2-0.85 keV and 0.85-12 keV energy bands. The red circles shows the locus of $R_{500}$.}
  \label{author1:fig}
\end{figure}

In the following we present the mock observation of a distant galaxy cluster (z=2, $M_{500}$ = 7 $10^{13}$ $M_\odot/h$) extracted from the numerical simulation of structure formation HYDRANGEA \citep{bahe}. These cosmological simulations are based on the EAGLE simulations, and use the same implementation of several physical processes for the intra-cluster medium (ICM). Each gas particles forming the ICM is considered as an X-ray source and its emission is modeled as a thermal emitting fully ionised plasma with emission lines (model \texttt{vvapec} in XSPEC \citep{arnaud}). The chemical abundances are in units of solar abundances as from \cite{anders}. In addition to the cluster X-ray emission, our mock observations include realistic background contributions as non X-ray background \citep{lotti}, cosmic X-ray background and foreground emission from the galaxy and the local bubble. Then, we make use of the end-to-end simulator SIXTE \citep{dauser} to produce an X-IFU event list. Following the methodology described in \cite{cucchetti}, we performed a $10^6$ sec exposure time mock observation (Fig.~\ref{author1:fig} middle).

\section{Reconstruction of the gas physical properties}

With this mock observation our goal is to demonstrate the ability of X-IFU to recover the physical properties of the ICM, that is the density, temperature, entropy, pressure. Making use of X-IFU's exquisite spectral resolution we also aim at estimating our ability to recover the  chemical abundances and abundance ratios.
Our analysis is starting. We first plan to recover the 1D projected distribution assuming the spherical symmetry of the cluster. In the second step we will look into the 2D reconstruction of the aforementioned physical characteristics. This requires an adequate radial or 2D binning (for instance through Voronoi tessellation) maximising the signal with respect to the objectives (e.g., measurements of temperature, chemical abundances). The challenges reside in the relatively modest number of cluster counts with respect to these of the total background (astrophysical and instrumental). Disentangling the line emission from the continuum is another difficulty. The X-IFU's exquisite spectral resolution and field of view will be a decisive asset in our study.

\section{Perspectives}

With this case study we will demonstrate the abilities of X-IFU to effectively observe and characterise the first massive haloes that have assembled an atmosphere of hot gas. The data analysis of our realistic  mock observations will allow us to quantify the precision with which we can recover the physical properties of the ICM. Hence, it will tells us with which precision we could constraint the amount of energy deposited in the ICM through feedback processes by star formation in galaxies and the activity of their super-massive black holes, together with, what is the chemical enrichment state of the first massive formed halos.

\begin{acknowledgements}
F. Castellani acknowledges the financial support of CNES (Centre National d'Etudes Spatiales) and ADS (Airbus Defence \& Space). We thank J. Schaye for his fruitful comments on this manuscript.
\end{acknowledgements}

\bibliographystyle{aa}  
\bibliography{Castellani_S17} 

\end{document}